\documentclass[12pt]{article}
\usepackage[english]{babel}
\usepackage{amsmath}
\usepackage{amsthm}
\usepackage{amsfonts}
\usepackage{amssymb}
\usepackage{euscript}
\usepackage[cp1251]{inputenc}
\usepackage[pdftex]{graphicx}
\newcommand{\comment}[1]{}
\usepackage[12pt]{extsizes}

\theoremstyle{definition}

\begin{document}
\begin{center}
  \textbf{Commuting differential operators of rank 2\\ with polynomial coefficients.}
\end{center}
\begin{center}
  \textbf{V.Oganesyan.}
\end{center}
\begin{center}
   \textbf{Introduction.}
\end{center}
If two differential operators
\begin{equation*}
L_n= \sum\limits^{n}_{i=0} u_i(x)\partial_x^i,  \quad  L_m= \sum\limits^{m}_{i=0} v_i(x)\partial_x^i
\end{equation*}
commute, then there is a nonzero polynomial $R(z,w)$ such that  $R(L_n,L_m)=0$ (see ~\cite{Chaundy}). The curve $\Gamma$ defined by $R(z,w)=0$ is called the \emph{spectral curve}. If\\
\begin{equation*}
L_n \psi=z\psi, \quad  L_m \psi=w\psi,
\end{equation*}
then $(z,w) \in \Gamma$. For almost all $(z,w) \in \Gamma$ the dimension of the space of common eigenfunctions $\psi$ is the same. The dimension of common eigenfunctions of two commuting differential operators is called the \emph{rank}. The rank is a common divisor of m and n.\\
If the rank equals 1, then there are explicit formulas for coefficients of commutative operators in terms of Riemann theta-functions (see ~\cite{theta}).\\
The case when rank is greater than one is much more difficult. The first examples of commuting ordinary scalar differential operators of the nontrivial ranks 2 and 3 and the nontrivial genus g=1 were constructed by Dixmier ~\cite{Dixmier} for the nonsingular elliptic spectral curve $w^2=z^3-\alpha$, where $\alpha$ is arbitrary nonzero constant:
\begin{equation*}
L= (\partial_x^2 + x^3 + \alpha)^2 + 2x ,
\end{equation*}
\begin{equation*}
M= (\partial_x^2 + x^3 + \alpha)^3 + 3x\partial_x^2 + 3\partial_x + 3x(x^2+\alpha) ,
\end{equation*}
where $L$ and $M$ is the commuting pair of the Dixmier operators of rank 2, genus 1. There is an example
\begin{equation*}
L= (\partial_x^3 + x^2 + \alpha)^2 + 2\partial_x  ,
\end{equation*}
\begin{equation*}
M= (\partial_x^3 + x^2 + \alpha)^3 + 3\partial_x^4 +  3(x^2+\alpha)\partial_x + 3x  ,
\end{equation*}
where $L$ and $M$ is the commuting pair of the Dixmier operators of rank 3, genus 1.\\
The general classification of commuting ordinary differential operators of rank greater than 1 was obtained by Krichever ~\cite{ringkrichever}. The general form of commuting operators of rank 2 for an arbitrary elliptic spectral curve was found by Krichever and Novikov ~\cite{novkrich}. The general form of operators of rank 3 for an arbitrary elliptic spectral curve (the general commuting operators of rank 3, genus 1 are parametrized by two arbitrary functions) was found by Mokhov ~\cite{Mokhov1},~\cite{Mokhov2}. Moreover, examples of commuting ordinary differential operators of arbitrary genus and arbitrary rank with polynomial coefficients were constructed in  ~\cite{Mokhov4}. However, even in those cases for which explicit formulas were obtained the task of allocating commuting operators with polynomial coefficients is non-trivial and not completely solved until now. The task of a complete description of commuting differential operators with polynomial coefficients was posed and discussed by Mokhov in ~\cite{Mokhov3}.\\
Mironov in ~\cite{Mironov} constructed examples of operators
\begin{equation*}
L = (\partial_x^2 + A_3x^3+ A_2x^2 + A_1x + A_0 )^2 + g(g+1)A_3x ,
\end{equation*}
\begin{equation*}
M^2 = L^{2g+1} + a_{2g}L^{2g} + ... + a_1L + a_0 ,
\end{equation*}
where $a_i$  are some constants and $A_i$, $A_3 \neq 0$, are arbitrary constants. Operators $L$ and $M$ are commuting operators of rank 2, genus g.\\
Furthermore, in ~\cite {Mironov2} it was proved by Mironov that the operators $L_1$ and $M_1$,
\begin{equation*}
L_1 = (\partial_x^2 + \alpha_1 \mathcal{P}(x) + \alpha_0)^2 + \alpha_1g_2g(g+1)\mathcal{P}(x), \quad \alpha_1 \neq 0 ,
\end{equation*}
\begin{equation*}
M_1^2 = L_1^{2g+1} + a_{2g}L_1^{2g} + ... + a_1L_1 + a_0 ,
\end{equation*}
where $a_i$  are some constants, $\alpha_i$ are arbitrary constants and $\mathcal{P}$ satisfies the equation
\begin{equation*}
(\mathcal{P'}(x))^2 = g_2\mathcal{P}^2(x) + g_1\mathcal{P}(x) + g_0,  \quad g_2 \neq 0,
\end{equation*}
where $g_1$ and $g_2$ are arbitrary constants, are a commuting pair of rank 2, genus $g$.\\
In the same paper it is proved that the operators $L_2$ and $M_2$,
\begin{equation*}
L_2 = (\partial_x^2 + \alpha_1 \wp(x) + \alpha_0)^2 + s_1 \wp(x) + s_2 \wp^2(x),
\end{equation*}
\begin{equation*}
M_2^2 = L_2^{2g+1} + b_{2g}L_2^{2g} + ... + b_1L_2 + b_0,
\end{equation*}
where $b_i$ are some constants, $\wp(x)$ is the Weierstrass elliptic function,\\  $\alpha_1=\frac{1}{4} - 2g^2 - 2g$,
$s_1=\frac{1}{4}g(g+1)(16\alpha_0 + 5g_2)$,  $s_2=-4g(g+2)(g^2-1)$,\\
 are also a commuting pair of rank 2, genus $g$.
\\
In this paper we find new examples of commuting operators of rank 2. The following theorems are proved\\
\\
\textbf{Theorem 1.} \emph{ Operator} $L=(\partial_x^2 + A_6x^6 + A_2x^2 )^2 + 16g(g+1)A_6x^4$, \emph{where} $g \in \mathbb{N}$,$A_6 \neq 0$ \emph{and} $A_2$ \emph{is an arbitrary constant}, \emph{commutes with a differential operator M of order} $4m+2$, \emph{for all} $m\geqslant g$. \emph{The spectral curve has the form} $w^2=z^{2m+1} + a_{2m}z^{2m} + ...+a_1z + a_0$. \emph{The operators} $L$ \emph{and} $M$ \emph{are operators of rank 2}.\\
\\
\textbf{Theorem 2.} \emph{ Operator} $L=(\partial_x^2 + A_4x^4 + A_2x^2 + A_0)^2 + 4g(g+1)A_4x^2$, \emph{where}  $g \in \mathbb{N}$, $A_4\neq 0$ \emph{and} $A_2, A_0$ \emph{are arbitrary constants}, \emph{commutes with a differential operator M of order} $4m+2$, \emph{for all} $m\geqslant g$. \emph{The spectral curve has the form} $w^2=z^{2m+1} + a_{2m}z^{2m} + ...+a_1z + a_0$. \emph{The operators} $L$ \emph{and} $M$ \emph{are operators of rank 2}.\\
\\
\textbf{Theorem 3.} \\
1) \emph{ If} $L=(\partial_x^2 + A_nx^n + A_{n-1}x^{n-1} + ... + A_0)^2 + B_kx^{k} + B_{k-1}x^{k-1} + ...+B_0$, \emph{where} $n>3, n \in \mathbb{N}$, $A_n \neq 0$, $B_k \neq 0$, \emph{commutes with an operator M of order} $4g+2$ \emph{and M, L are operators of rank 2}, \emph{then} $k=n-2$ \emph{and} $B=(n-2)^2m(m+1)A_n$ \emph{for some} $m \in \mathbb{N}$.\\
2) \emph{For} $n>6$ \emph{ operator $L=(\partial_x^2 + A_nx^n)^2 + B_{n-2}x^{n-2}$ does not commute with any differential operator} $M$ \emph{of order} $4g+2$, \emph{where} $M, L$ \emph{could be a pair of rank 2}.\\
3) \emph{ If} $n=5$, \emph{then} $L=(\partial_x^2 + Ax^5)^2 + 18Ax^{3}$, $A\neq 0$, \emph{ commutes with an operator M of order} $4g+2$ \emph{for all} $g$ \emph{and M,L are operators of rank 2}. \\
4) \emph{Operator}  $(\partial_x^2 + Ax^5)^2 + 9m(m+1)Ax^{3}$, $A\neq 0$ \emph{does not commute with any differential operator M of order} $4g+2,$ \emph{for} $m>1$, \emph{where M and L could be a pair of rank 2} .\\
\\
Note that for genus less than 9, for $ m = g $, the spectral curve of the operator from theorem 1 is non-singular for almost all $A_2$. Apparently, when $ m = g $, the spectral curve is non-singular for any  $g$, but it is not proved. For $m>g$, apparently, the spectral curve is always singular. Here are few examples. For $m=g=1$ the spectral curve of operator from theorem 1 has the form
\begin{equation*}
w^2=(z + 16A_2)(z^2+ 16A_2z +192A_6).
\end{equation*}
For $g=1, m=3$, we have
\begin{equation*}
w^2=(z + 16A_2)(z^2+ 16A_2z +192A_6)(256A_2^2 + C_2 - 16A_2C_1-zC_1(16A_2 - 1) + z^2)^2,
\end{equation*}
where $C_1$, $C_2$, $C_3$ are arbitrary constants. For $m=g=3$ and $A_2=0$
\begin{equation*}
w^2=z(z^2+288000A_6)(273715200A^2_6 + 289152A_6z^2 + z^4).
\end{equation*}
The operator from theorem 2 has singular spectral curve for $m=g=2,5$. Apparently the spectral curve is singular only for $m=g=3k-1$. Here are  few examples. For $m=g=1$ the spectral curve has the form
\begin{equation*}
w^2=z^3 + 8A_2z^2 + 16(A_2^2 + A_0A_4)z + 64A_0A_2A_4 + 16A_4^2.
\end{equation*}
For $g=1, m=3$ and $A_2=A_0=0$
\begin{equation*}
w^2=(z^3 + 16A_4^2)(C_2 + C_1z + z^2)^2,
\end{equation*}
where $C_1,C_2$  are arbitrary constants. If $g=m=3$ and $A_2=A_0=0$
\begin{equation*}
w^2=z(3382560000A_4^4 + 117216A_4^2z^3 + z^6).
\end{equation*}
Theorem  3 is not true for the general form of polynomials\\ $Ax^n+A_{n-1}x^{n-1} + ...+A_1x+A_0$ and $Bx^{n-2}+B_{n-1}x^{n-1}+...+B_1x+B_0$. For example, for any $n$ there is a differential operator of order 6  that commutes with $L$. There is a reason to believe that $L$ commutes with differential operator of order $4g +2$ for $g>1$ only for $ n = 3,4,5,6 $, but it is not proved.\\
Note that the spectral curve of operator $(\partial_x^2 + Ax^5)^2 + 18Ax^{3}$ is always singular.\\
\\
The author is grateful to O.I.Mokhov for valuable discussions.
\begin{center}
   \textbf{Commuting operators of rank 2.}
\end{center}
Consider the operator
\begin{equation}
(\partial^2_x + V(x))^2 + W(x)
\end{equation}
From article \cite{Mironov} we know that the operator  commutes with an operator of order $4g+2$ with hyperlyptic spectral curve of genus $g$ and hence is operator of rank 2, if and only if there is a polynomial
\begin{equation*}
Q=z^g + a_1(x)z^{g-1} + a_2(x)z^{g-2} + ...+a_{g-1}(x)z + a_g(x)
\end{equation*}
that the following relation is satisfied
\begin{equation}
Q^{(5)} + 4VQ''' + 6V'Q'' + 2Q'(2z-2W+V'') - 2QW'\equiv 0,
\end{equation}
$Q'$ means $\partial_xQ$. The spectral curve has the form
\begin{equation}
4w^2=4F(z)=4(z-W)Q^2 - 4V(Q')^2 + (Q'')^2 - 2Q'Q''' + 2Q(2V'Q' + 4VQ'' + Q^{(4)}).
\end{equation}
Using (2), we get
\begin{equation*}
4Q'z\equiv -Q^{(5)} - 4VQ''' - 6V'Q'' - 2Q'V'' + 2QW' + 4Q'W.
\end{equation*}
So we have the following system :
\begin{equation*}
\left\{
 \begin{array}{l}
a_1=W/2 + C_1\\
4a_2' = -a_1^{(5)} - 4Va_1''' - 6V'a_1'' - 2a_1'V'' + 2a_1W' + 4a_1'W\\
...\\
4a_{i+1}' = -a_i^{(5)} - 4Va_i''' - 6V'a_i'' - 2a_i'V'' + 2a_iW' + 4a_i'W\\
...\\
4a_{g}' = -a_{g-1}^{(5)} - 4Va_{g-1}''' - 6V'a_{g-1}'' - 2a_{g-1}'V'' + 2a_{g-1}W' + 4a_{g-1}'W\\
0 = -a_g^{(5)} - 4Va_g''' - 6V'a_g'' - 2a_g'V'' + 2a_gW' + 4a_g'W
 \end{array}
\right.  \label{sys}
\end{equation*}
It is clear from the system that result in \cite{Mironov} can be formulated in the following way. Let $a_1=W/2 + C_1$, where $C_1$ is an arbitrary constant. Define $a_i$ by recursion
\begin{equation}
a_{i+1}=\frac{1}{4}\int(-a_i^{(5)} - 4Va_i''' - 6V'a_i'' - 2a_i'V'' + 2a_iW' + 4a_i'W)dx
\end{equation}
We see that (1) commutes with an operator of order $4g+2$ and these operators are operators of rank 2 if and only if $a_{g+1}\equiv const$. For example,
\begin{equation*}
a_{2}=-\frac{1}{8} W^{(4)} - \frac{1}{2}VW'' - \frac{1}{4}V'W' + \frac{3}{8}W^2 + \frac{1}{2}WC_1 + 2C_2 ,
\end{equation*}
where $C_2$ is a constant, which appears after integration. In general, $a_i$ contains $i$ constants, i.e., $a_i(x)=a_i(x;C_1,...,C_i)$.\\
\\
\\
\begin{center}
   \textbf{Proof of Theorem 1.}
\end{center}
Suppose that $V=A_6x^6 + A_2x^2  $, $W=16A_6g(g+1)x^4$, where $g \in \mathbb{N}$. The operator has the form
\begin{equation}
(\partial^2_x + A_6x^6 + A_2x^2)^2 + 16A_6g(g+1)x^4.
\end{equation}
Let us prove that we can choose constants $C_1,C_2,...,C_g$ in such a way that  $a_{g+1}=const$. Calculation shows that
\begin{equation*}
a_1= C_1 + 8A_6g(g+1)x^4.
\end{equation*}
It is clear from (4) that $a_{i+1}$ is polynomial in $x$ and linear in $a_i$.
Let us apply (4) to  $x^{4k}$. Assume that $a_i=x^{4k}$\\

$a_{i+1}= C_{i+1} - k(4k-1)(4k-2)(4k-3)x^{4k-4} - 16A_2k^2x^{4k}$\\$+ \frac{8A_6}{k+1}(2k+1)(g-k)(g+k+1)x^{4k+4} $.\\
\\
Using this formula we can calculate the following
\begin{equation*}
a_2=\widetilde{C}_2 + 8A_6(\widetilde{C}_1 - 16A_2)g(g+1)x^4 +  96A^2_6g(g+1)(g-1)(g+2)x^8,
\end{equation*}
\begin{equation*}
a_3=\widetilde{C_3} + 8A_6g(g+1)(256A_2^2 - 16A_2\widetilde{C_1} + \widetilde{C_2} - 5040A_6(g-1)(g+2))x^4 -
\end{equation*}
\begin{equation*}
-96A_6^2(80A_2-\widetilde{C}_1)g(g+1)(g-1)(g+2)x^8+1280A^3_6g(g+1)(g-1)(g+2)(g-2)(g+3)x^{12},
\end{equation*}
where $\widetilde{C}_i$ are constants and depend on $C_i$. During the transition from $a_3$ to $a_4$ the term $\widetilde{C}_3$   will be multiplied by $8A_6g(g+1)x^4$. Also $16A_2\cdot const \cdot x^4$ and the fourth derivation of $const \cdot x^8$ will be subtracted from $8\widetilde{C}_3A_6g(g+1)x^4$. Polynomial $a_4$ will have constant term $\widetilde{C}_4$.\\
\\
Generally, if $a_{i}=\widetilde{C_i} +  K_{i}^1x^4 + K^2_{i}x^{8} + ...+K^i_{i}x^{4i}$, then \\
\\
$a_{i+1}=\widetilde{C}_{i+1} + (8A_6g(g+1)\widetilde{C_i} - 16A_2K^1_i - 2\cdot 7\cdot 6\cdot 5 K^2_i)x^4 + $\\
\\
$+(const\cdot K^1_i - const\cdot K^2_i - const \cdot K^3_{i})x^{8}  + .. +\frac{8A_6(2i+1)}{i+1}(g-i)(g+i+1)K^i_{i}x^{4i+4}$.\\
\\
Also note that all coefficients, except the leading term, contain constants $C_1,C_2,....$ And $K^{i-1}_{i}$ contains only $C_1$, $K^{i-2}_{i}$ contains only $C_1$ and $C_2$,  $K^{i-3}_{i}$ only $C_1, C_2, C_3$ etc.\\
We see that the term $const \cdot x^{4(g+1)}$ in $a_{g+1}$ vanishes because it is multiplied by $const\cdot (g-g)(g+g+1)=0$.
Therefore, we must choose the constants $C_1, C_2,...$ to vanish $K_{g+1}^m=0$ for all $m$. This is always possible because the leading term in $a_{g+1}$ equals $const \cdot x^{4g}$ and depends only on $C_1$, penultimate on $C_1$ and $C_2$ etc.\\
Let us suppose we didn't choose constants to vanish $a'_{g+1}$ and continued applying transformation (4). Consider $a_{g+1+m}$, $m>0$. Note that $a_{g+1+m}$ has the same structure as $a_{g+1}$ with the only difference that $a_{g+1+m}$ contains constants $C_1, C_2,...,C_{g+m}, C_{g+1+m}$. This means that we can choose constants $C_1,...,C_{g+1+m}$ in such a way that $a'_{g+1+m} \equiv 0$. Hence there is a differential operator $M$ of order $4(g+m) + 2$ such that $M$ commutes with $L$ and they are operators of rank 2.
\begin{center}
    \textbf{Proof of Theorem 2}.
\end{center}
The operator has the form
\begin{equation}
(\partial^2_x + A_4x^4 + A_2x^2 + A_0)^2 + 4g(g+1)A_4x^2.
\end{equation}
As in the previous proof, we will prove that constants $C_1,C_2,...,C_g$ can be chosen in such a way that  $a_{g+1}=const$. Calculation shows that
\begin{equation*}
a_1= C_1 + 2A_4g(g+1)x^2 .
\end{equation*}
We mentioned before that $a_{i+1}$ is polynomial in $x$ and linear in $a_i$.
Let us apply (4) to  $x^{2k}$. Assume that $a_i=x^{2k}$, then  we have\\
\\
$a_{i+1}= C_{i+1} -k(2k-1)(k-1)(2k-3)x^{2k-4} - 2A_0k(2k-1)x^{2k-2} - $\\\\$- 4A_2k^2x^{2k} + \frac{2A_4(2k+1)(g-k)(g+k+1)x^{2k+2}}{k+1} $.\\
\\
Using this formula we can calculate the following
\begin{equation*}
a_2=\widetilde{C_2} + 2A_4(\widetilde{C}_1 - 4A_2)g(g+1)x^2  +6A^2_4g(g+1)(g-1)(g+2)x^4,
\end{equation*}
\begin{equation*}
a_3= \widetilde{C_3} + 2A_4g(g+1)(16A_2^2 - 4A_2\widetilde{C}_1 + \widetilde{C_2} - 36A_0A_4(g-1)(g+2))x^2 -
\end{equation*}
\begin{equation*}
-6A_4^2(20A_2 - \widetilde{C}_1)g(g+1)(g-1)(g+2)x^4  +  20A^3_4g(g+1)(g-1)(g+2)(g-2)(g+3)x^{6}
\end{equation*}
where $\widetilde{C}_i$ are constants and depend on $C_i$.\\
\\
Generally, let $a_{i}=\widetilde{C}_i +  K_{i}^1x^2 + K^2_{i}x^{4} + ...+K^i_{i}x^{2i}$, then \\
\\
$a_{i+1}=\widetilde{C}_{i+1} + (2A_4g(g+1)\widetilde{C}_{i} - 4A_2K_i^1 -12A_0K_i^2 - 90K^3_i)x^2 + $\\
\\
$+(const\cdot K_{i}^1 - const \cdot K^2_{i} - const \cdot K^3_{i} - const \cdot K^4_{i})x^{4}  + .. +$\\
\\
$+ \frac{2A_4(2i+1)}{i+1}(g-i)(g+i+1)K^i_{i}x^{2i+2}$.\\
\\
As in the proof of the previous theorem all terms except the leading term contain constants $C_1,C_2,...$. And $K^{i-1}_{i}$ contains only $C_1$, $K^{i-2}_{i}$ contains only $C_1,C_2$ and $K^{i-3}_{i}$ only $C_1, C_2, C_3$ etc.\\
The term $const \cdot x^{2(g+1)}$ in $a_{g+1}$ vanishes because it is multiplied by\\ $\frac{2A(2g+1)}{g+1}(g-g)(g+g+1)=0$. So we must choose constants $C_1, C_2,...$ in such a way that $A_{g+1}^i=0$ for all $i$. This is always possible because the leading term in $a_{g+1}$ depends only on $C_1$, penultimate on $C_1$ and $C_2$ etc.\\
Let us suppose we didn't choose constants to vanish $a'_{g+1}$ and continued applying transformation (4). Consider $a_{g+1+m}$, $m>0$. Note that $a_{g+1+m}$ has the same structure as $a_{g+1}$ with the only difference that $a_{g+1+m}$ contains constants $C_1, C_2,...,C_{g+m}, C_{g+1+m}$. This means that we can choose constants $C_1,...,C_{g+1+m}$ in such a way that $a'_{g+1+m} \equiv 0$. Hence there is a differential operator $M$ of order $4(g+m) + 2$ such that $L$ commutes with M.
\begin{center}
    \textbf{Proof of Theorem 3}.
\end{center}
If $L$ commutes with operator of order $4g+2$, then $a_{g+1} \equiv 0.$ We see that $a_i(x)$ is polynomial in $x$. Assume that degree of $a_g$ equals $l$. Then degree of $-4Va_g''' - 6V'a_g'' - 2a'_gV''$ is equal to $n+l-3$. Sum $-4Va_i''' - 6V'a_i'' - 2a_i'V'' \neq 0$ because the leading coefficient is not equal to zero. Because of the same reason $2a_gW' + 4a_g'W \neq 0$. The degree of $2a_gW' + 4a_g'W$ equals $k+l-1$. So, if $a_{g+1}\equiv 0$, then $n+l-3=k+l-1$, $\Rightarrow$ $k=n-2$.\\
Let us prove that $B_{n-2}=(n-2)^2m(m+1)A_n$ for some $m \in \mathbb{N}$. We will consider only leading coefficient of $a_i(x)$ and without loss of generality it can be assumed that the operator has the form
\begin{equation}
L=(\partial^2_x + A_nx^n )^2 + B_{n-2}x^{n-2},
\end{equation}
We get
\begin{equation*}
a_1= C_1 + \frac{1}{2}B_{n-2}x^{n-2}.
\end{equation*}
We know that $a_{i+1}$ is polynomial in $x$ and linear in $a_i$.
Let us apply (4) to  $x^{k}$. Assuming that $a_i=x^{k}$
\begin{equation}
a_{i+1}=-\frac{1}{4}k(k-1)(k-2)(k-3)x^{k-4} +
\end{equation}
\begin{equation*}
+\frac{(n+2k-2)}{2(n+k-2)}(B_{n-2}-Ak(n+k-2))x^{n+k-2} + C_{i+1}.
\end{equation*}
We see from this formula that the leading term of $a_i$ is  term of degree $(n-2)i$.\\
If $a'_{g+1} \equiv 0$,  then for some  $m\leqslant g$  the leading term of  $a_{m+1}$ must vanish. Assuming in (4) that $a_i=x^{i(n-2)}$
\begin{equation*}
a_{i+1}=-\frac{1}{4}i(n-2)(i(n-2)-1)(i(n-2)-2)(i(n-2)-3)x^{i(n-2)-4} +
\end{equation*}
\begin{equation*}
+ \frac{(2i+1)}{2(i+1)}(B_{n-2}-Ai(i+1)(n-2)^2)x^{i(n-2) + n-2} +  C_{i+1}.
\end{equation*}
From this formula we have
\begin{equation*}
B_{n-2}=(n-2)^2m(m+1)A_n.
\end{equation*}
This means that we can rewrite (8) as
\begin{equation}
-\frac{1}{4}k(k-1)(k-2)(k-3)x^{k-4} +
\end{equation}
\begin{equation*}
+\frac{A_n(n+2k-2)}{2(n+k-2)}(m(n-2)-k)((m+1)(n-2)+k)x^{n+k-2} + C_{i+1}.
\end{equation*}
\\
Let us prove items 2--4. Assume $n>6$.\\
Coefficient of the leading term in $a_i$ is positive for $i<m$ because we see  from (9) that during the transition from $a_{i}$ to $a_{i+1}$ the leading term will be multiplied by
\begin{equation*}
\frac{A(2i+1)}{2(i+1)}(n-2)^2(m-i)(m+i+1)x^{n-2}.
\end{equation*}
By $N_k$ denote $\frac{(n+2k-2)}{2(n+k-2)}(m(n-2)-k)((m+1)(n-2)+k)$.\\
As it was mentioned before $N_{i(n-2)}>0$ for $i<m$. Easy to check that $N_{i(n-2)-4}>0$ for $i\leqslant m$, $i \geqslant 1$. Using (9), we get\\
\begin{equation*}
a_2 =\widetilde{C}_2 - \frac{1}{8}Am(m+1)(n-5)(n-4)(n-3)(n-2)^3x^{n-6} + const \cdot x^{n-2} +
\end{equation*}
\begin{equation*}
+\frac{3}{8}A^2(m-1)m(m+1)(m+2)(n-2)^4x^{2n-4}.
\end{equation*}
Note that the coefficient of the term $x^{n-6}$ is negative. It is clear from (9) that during the transition to $a_3$ the term $A\cdot const \cdot x^{n-6}$ will be multiplied by $AN_{n-6} x^{n-2}=AN_{(n-2)-4}x^{n-2}$ and the fourth derivation of $A^2\cdot const \cdot x^{2n-4}$ will be subtracted from $A\cdot const \cdot x^{2n-8}$. This means that the coefficient of $x^{2n-8}$ in $a_3$ equals  $A^2\cdot const\cdot x^{2n-8}$, where constant is a negative number. So if $i\leqslant m$, then during the transition from $a_i$  to $a_{i+1}$ the coefficient of $x^{(i-1)(n-2)-4}$ will be multiplied by $AN_{(i-1)(n-2)-4}x^{n-2}$ and a positive number will be subtracted (the fourth derivation of $const\cdot x^{i(n-2)}).$\\
We see that if $i\leqslant m$,  then  the term $A^{i-1}\cdot const \cdot x^{(i-1)(n-2)-4}$ vanishes only if $A=0$ because $const<0$.\\
During the transition from $a_m$ to $a_{m+1}$ the leading term $const \cdot x^{m(n-2)}$ will be multiplied by $0$ and will vanish.\\
If $i> m$, then during the transition from $a_i$ to $a_{i+1}$ nothing will be subtracted from the term $A^{i-1}{\cdot}const\cdot x^{(i-1)(n-2)-4}$ . We see from (9) that during the transition to $a_{i+1}$ the member $x^{(i-1)(n-2)-4}$ can vanish only because of factor $m(n-2) - (i-1)(n-2)+4$. Suppose $i-m \geqslant 2$. Before we assumed that $n>6$.
\begin{equation*}
m(n-2) - (i-1)(n-2)+4=(n-2)(m-i+1)+4 \leqslant -(n-2) + 4 < 0.
\end{equation*}
When $i-m=1$
\begin{equation*}
m(n-2) - (i-1)(n-2)+4=4.
\end{equation*}
Finally, we obtain that if $n>6$ and $A \neq 0$, then there is no such  $g$ that $a'_{g+1} \equiv 0$.\\\\
Assume $n=5$. Consider operator
\begin{equation}
L=(\partial^2_x + Ax^5)^2 + 18Ax^3.
\end{equation}
Calculations show that
\begin{equation*}
a_2 =\widetilde{C}_2 +9A\widetilde{C}_1x^3 .
\end{equation*}
So if $\widetilde{C}_1=0$, then $a'_2 \equiv 0$. Hence, there is an operator $M$ of order 6 such that M commutes with $L$.
\begin{equation*}
a_3 =\widetilde{C}_3 +9A\widetilde{C}_2x^3
\end{equation*}
If $\widetilde{C}_2=0$, then there is an operator $M$ of order 10 such that M commutes with $L$. In general
\begin{equation*}
a_{g+1} =\widetilde{C}_{g+1} +9A\widetilde{C}_gx^3.
\end{equation*}
If $\widetilde{C}_g=0$, then there is an operator M of order $4g+2$ and M commutes with $L$.\\
\\
Let $L=(\partial^2_x + Ax^5)^2 + 9m(m+1)Ax^3$, where $m>1$. Then
\begin{equation*}
a_{2} =\widetilde{C}_2 + 9\widetilde{C}_1m(m+1)x^3 + \frac{243}{2}A^2m(m+1)(m+2)(m-1)x^6
\end{equation*}
\begin{equation*}
a_{3} =\widetilde{C}_3 - \frac{10935}{4}A^2m(m+1)(m+2)(m-1)x^2 + const\cdot x^3 + const\cdot x^6 + A^3\cdot const\cdot x^9
\end{equation*}
Arguing as before we will come to conclusion that the degree of  $x^2$ will grow and never vanish.
\\

Department of Geometry and Topology, Faculty of Mechanics and Mathematics, Lomonosov Moscow State University, Moscow, 119991 Russia.\\\\
E-mail address: vardan.o@mail.ru

\end{document}